# Dependence of Carbon Nanotube Field Effect Transistors Performance on Doping Level of Channel at Different Diameters: on/off current ratio

Shaahin G. Shirazi,[1,a)] and Sattar Mirzakuchaki[2]

[1]*Electrical, Computer & IT College of Islamic Azad University, Qazvin Branch, 34185-1416, Qazvin, Iran*

[2]*E.E. Department of Iran University of Science and Technology, Narmak, 13114-16846, Tehran, Iran*

Choosing a suitable doping level of channel relevant to channel diameter is considered for determining the carbon nanotube field effect transistors' performance which seem to be the best substitute of current transistor technology. For low diameter values of channel the ratio of on/off current declines by increasing the doping level. But for higher diameter values there is an optimum point of doping level in obtaining the highest on/off current ratio. For further verification, the variations of performance are justified by electron distribution function's changes on energy band diagram of these devices. The results are compared at two different gate fields.

Carbon nanotube (CNT) has promising electrical properties[1-3] to be used as a channel in field effect transistors (FETs). Carbon nanotube FETs (CNTFETs)[4,5] are devices in which CNT's superior electronics behavior results in very attractive properties[6]. The first CNTFET[4] was reported in 1998. Consequently their structures have improved and some amendments in their performance have occurred[7-10]. Recent CNTFET devices with extended source/drain doped region show lower leakage, high on-state current at ballistic regime compared to CNT Schottky

---

a) shaahin.shirazi@ymail.com

barrier FETs (CNT SBFETs)[11]. Such devices work like metal-oxide-semiconducting FETs (MOSFETs), and are called CNT MOSFETs. In this work, a CNT MOSFET will be studied. There is no significant research on these devices' performance based on changes in the doping level of channel at different diameters. Here, the on/off current ($I_{on}/I_{off}$) ratio is surveyed. The results are interesting and help us to choose optimum doping level of channel at a specific diameter to gain the highest performance. The performance changes are justified by considering the variations of electron distribution function of the device on its energy band diagram. A schematic of this device and its characteristic is represented in Figs. 1(a) and 1(b) in detail.

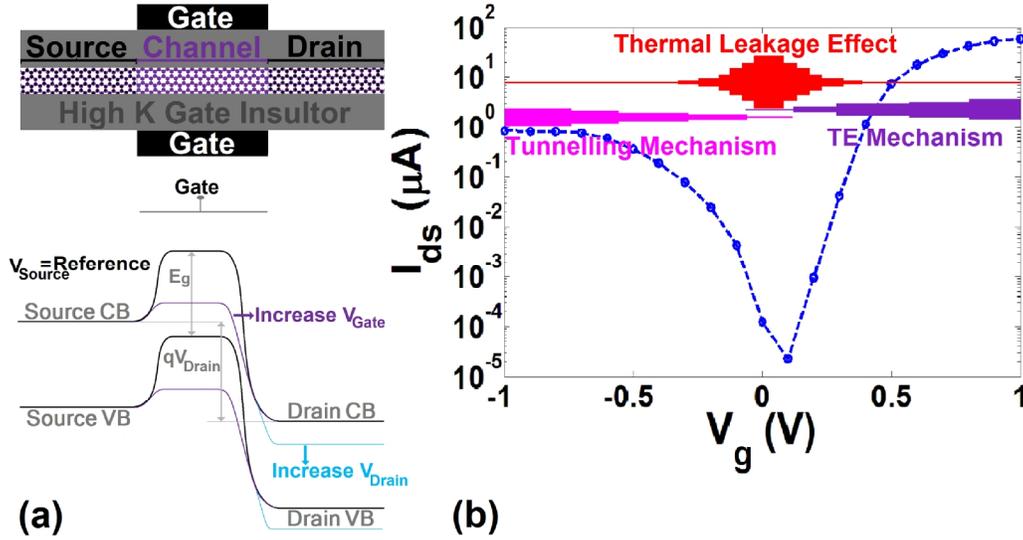

FIG. 1. (Color online) (a) Conventional coaxially-gated CNT MOSFET structure and its band diagram. Source/Channel junction height at gate field ($V_g$)=0V is $E_g/2$, and (b) current-voltage ($I_{ds}$-$V_g$) characteristic of a CNT MOSFET, drain field ($V_d$)=0.4V, gate length=15nm, source (or drain) length=30nm, (13,0) zigzag CNT channel, source/drain doping level=15×10$^8$m$^{-1}$, channel doping level ($N_c$)=0m$^{-1}$, gate insulator dielectric constant=25 and gate insulator thickness=2nm at room temperature.

The theoretical results extracted from bottom-up simulation approach show a well agreement with experimental results[11,12]. The best approach for simulating the transmission behavior of carriers in such devices at nanoscale is the non-equilibrium Green's function (NEGF) formalism[13-17] which solves Poisson and Schrödinger equations self consistently. Because of short length of the channel, i.e. 15nm, the transmission of carriers is supposed to be at ballistic regime.

In CNT MOSFETs there are two main mechanisms in the flow of the electrons from source to drain. The drain/source current versus the gate field ($I_{ds}$-$V_g$) characteristic of this transistor is shown in Fig. 1(b). In order to better understand the current mechanisms, applet tools and marvelous sources in nanoHUB.org[18-20] are used which help us to prove our idea and to explain the changes in the results using distribution function of electrons on energy band diagram of these devices. In Fig. 2, the electron distribution function of the device is illustrated for three different gate fields. According to Fig. 2(a), a high gate field pulls down the energy band of channel. Now the carriers can easily transmit from source to drain via thermionic emission (TE) mechanism and the value of current becomes high. When the gate field abates to lower fields, i.e. 0.4V (Fig. 2(b)), the height of the potential barrier between source and channel becomes smaller. Consequently the TE mechanism becomes less important and a second current mechanism will become important gradually. Valence band edge of channel will be closer to conduction band of source/drain and consequently tunneling states will appear in the channel (Fig 2(c)). This mechanism is called band-to-band tunneling. As is shown in Fig. 2(d), if the field is decreased further, because of an increase in the height of potential barrier, the easy transport of electron from source to drain, i.e. TE mechanism, absolutely degrades and the band-to-band tunneling mechanism becomes more significant. Also, leakage current which includes any current mechanism at zero gate field, becomes significant when both of TE and band-to-band tunneling mechanisms have a small value. The leakage current is very important in determining the device performance at the subthreshold region. The thermal component of leakage current is triggered by the electrons movement from source region across the top of the potential barrier. This is due to the lattice temperature which exists in all of the transmission stages. Since the leakage value is very small, it doesn't affect the high current values of the saturation and inverse saturation current.

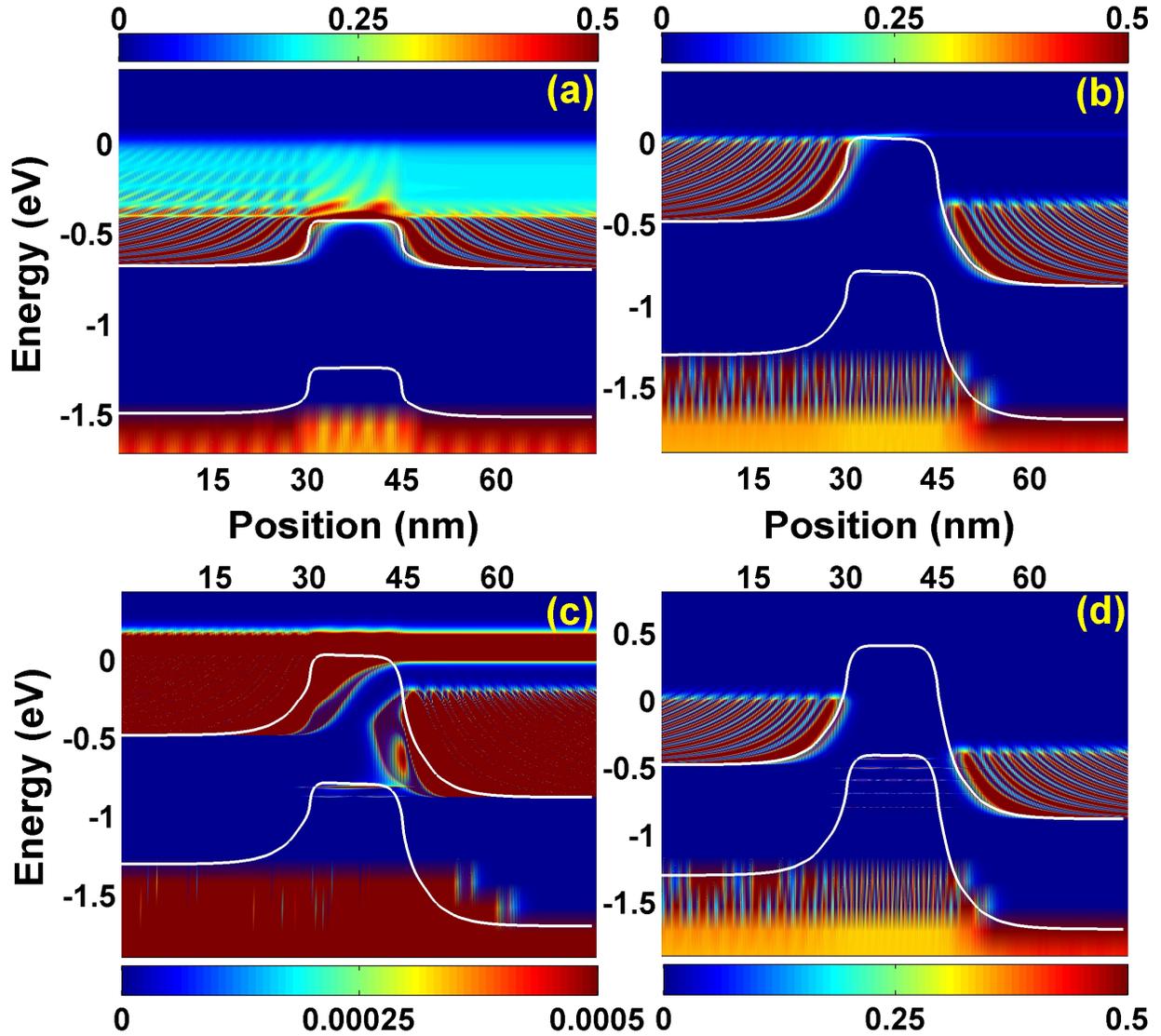

FIG. 2. (Color online) Electron distribution function of MOS CNTFET with the same conditions in Fig.1 at: (a) $V_g$=1V, (b) $V_g$=0.4V, (c) $V_g$=0.4V (1000 times more sensitive than case (b)), and (d) $V_g$=0V. In all cases maximum value of colormap band shows full of electrons regions and 0 (minimum value) shows empty of carriers regions.

Now the impact of the channel doping level on $I_{on}/I_{off}$ ratio of the device can be investigated in detail. If on-state gate field is supposed to be at 1V and off-state to be at 0V, it can be seen for low "n" values, e.g. n=10, that increasing the channel doping causes the current ratio to decrease continuously (Fig. 3(a)). When the position of electron distribution function for different "n" values is studied, it can be found that by increasing the doping of the channel, the height of potential barrier becomes lower (Fig. 4). The Fermi level of CNT channel with no doping is exactly at the middle of the band gap[1-3]. Also by raising the doping value, the Fermi level becomes closer to the conduction band and consequently the barrier height between source and channel decreases. It should be mentioned that in equilibrium the Fermi level of the CNT channel without doping is along the Fermi level of the source which is held at zero eV as a reference. By increasing the doping level of the channel, its Fermi level should also increase in the energy band diagram. On the other hand, the Fermi level is supposed to stay along the Fermi level of the source. It means that the energy band diagram of channel is pushed down. So the barrier height between the source and the channel decreases. As a result, although the tunneling states disappear gradually but it should be noted that thermal leakage current caused by the flow of carriers over top of the barrier at zero gate field, grows up. This is due to the negligible height of the potential barrier. This phenomenon is dominant compared with the reduction in tunneling leakage current. So the overall effect of increasing the doping level of channel for low "n" values is degradation in $I_{on}/I_{off}$ ratio. There is an inverse relation between the channel diameter, i.e. "n" value of channel, and its energy band gap[2,3]. Consequently for high "n" values, the valence band edge of channel is closer to the conduction band edge of the source at the same gate field than lower "n" values. For n=13, the tunneling states in such situation are more intense and the dependency of leakage current to the tunneling is more than in the case of n=10. For this reason, in the case of n=13 an increase in the doping level of the channel will result in a reduction in the tunneling current which is dominant in comparison with the intensification of thermal leakage current. This is in contrast to the case of n=10, and the final result is an improvement in the $I_{on}/I_{off}$ ratio. Consequently for n=10, the $I_{on}/I_{off}$ ratio curve in Fig. 3(a) has a downward trend. But for n=13, after a certain doping level of the channel, the tunneling states reduce dramatically and thermal leakage current causes the current ratio to degrade. Such event occurs for "n" value equal to 17. But at this value, because of smaller distance between valence and conduction band of channel compared to the previous cases of n=10 and n=13, the tunneling current is very dominant and therefore the increasing in current ratio continues up to $5\times10^8$ m$^{-1}$ doping level value.

Now, if we suppose on-state current at 0.4V gate field, corresponding to the 2010 International Technology Roadmap for Semiconductors (ITRS), as is shown in Fig. 3(a), there are some differences than 1V gate field. If $V_g$=1V, TE mechanism is the most dominant mechanism for transmission of carriers from source to drain and the on-state current of all cases remains almost constant. Consequently, in investigating the $I_{on}/I_{off}$ ratio only the changes in off-state current were needed. But at 0.4V gate field the current is composed of both TE and band-to-band tunneling components (Fig. 2(c)). This shows that the current ratio is not only dependent on the off-state but also on the on-state current. In this situation there is a key point to be considered. The tunneling states in such field are not strong and intense for almost all of "n" values. Then by increasing the channel doping level, the tunneling vanishes gradually but the TE component of current grows up because of a reduction in the height of potential barrier. Consequently, the on-state current always increases at any "n" values, while the current at zero gate field is the same as mentioned before. Therefore at 0.4V field, the rate of improvement in $I_{on}/I_{off}$ ratio is intensified compared to the 1V gate field although the overall value is smaller. Now, the details of these curves are explained. For low "n" values, e.g. n=10, we see at first an increase in ratio as the doping increases. This is opposite to the previous gate field of 1V where the curve had absolutely downward shape. It's because of on-state current impact on the ratio. At these steps while the challenge between decreasing the tunneling states at zero gate field and increasing thermal leakage leads to growth in leakage current, but increasing the TE current at on-state is more effective and therefore the $I_{on}/I_{off}$ ratio improves. But after a certain doping level, the impact of leakage at zero gate field, becomes dominant and the whole effect is a reduction in current ratio. Since the distance between valence and conduction band of channel at low "n" values is high, then the effect of tunneling and thermal leakage (due to higher potential barrier) are insignificant and consequently the $I_{on}/I_{off}$ ratio is very high. By further addition of doping in the channel, and hence a reduction in tunneling states, the performance doesn't change sharply at first. But for higher "n" values since the tunneling states become significant, the increase in doping, decreases the tunneling more and more and consequently the range of increase in ratio is more than lower "n" values up to a certain channel doping level amount.

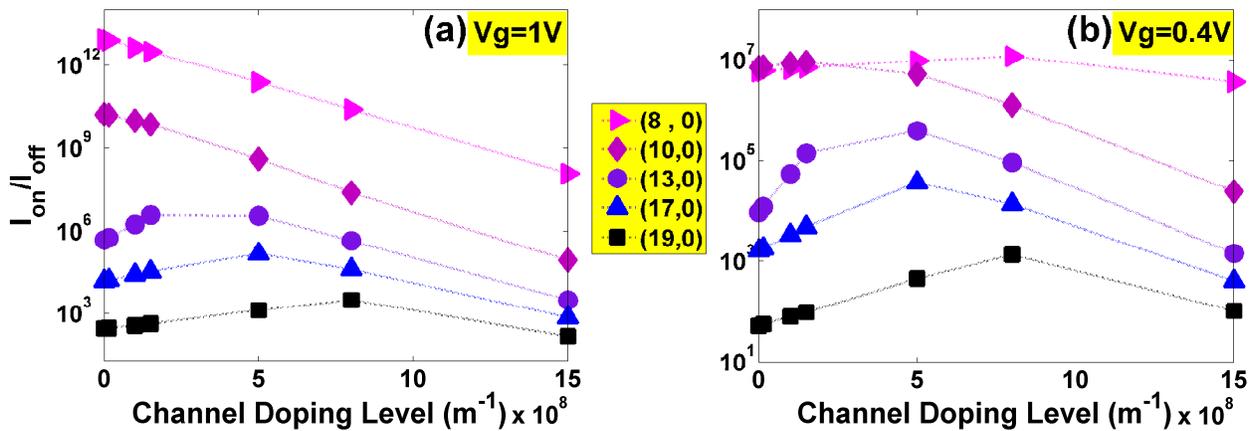

FIG. 3. (Color online) $I_{on}/I_{off}$ ratio vs. channel doping level for transistor with the same condition in Fig. 1 at different channel diameters at: (a) $V_g$=1V, and (b) $V_g$=0.4V gate fields.

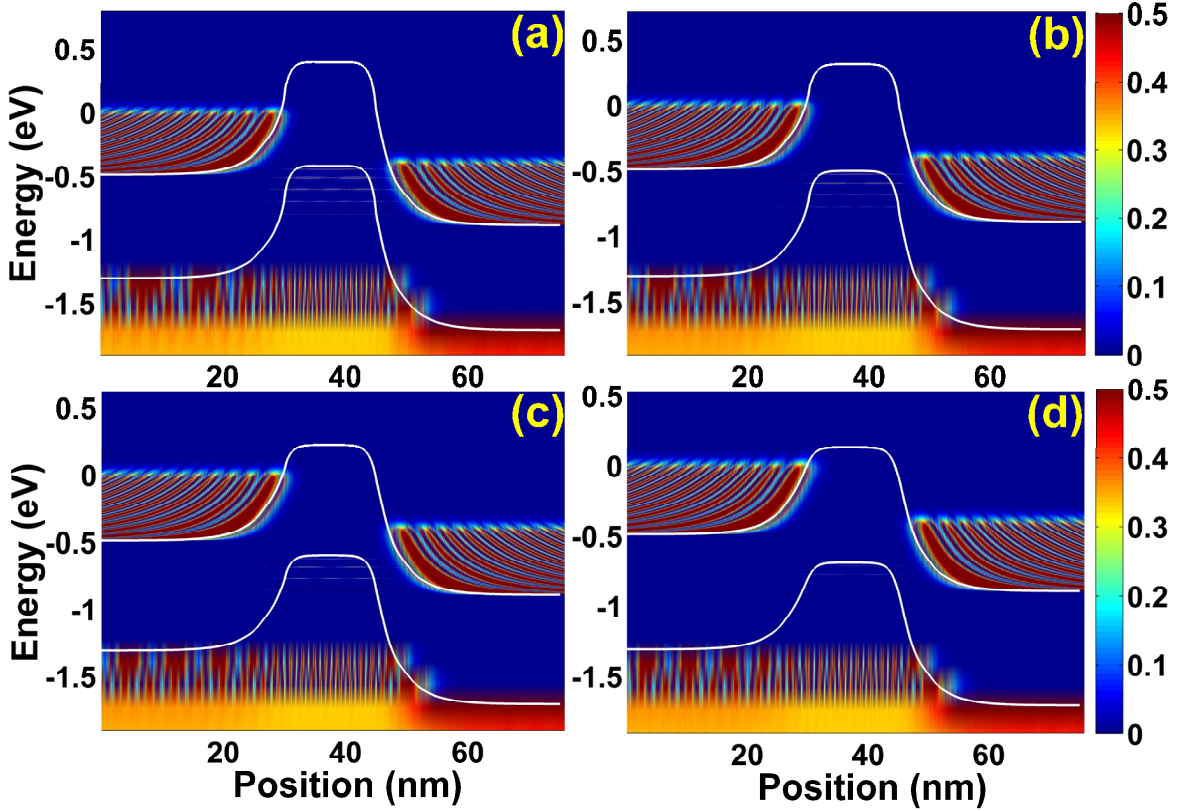

FIG. 4. (Color online) Electron distribution function of MOS CNTFET with the same conditions in Fig.1 at: (a) $N_c=0 m^{-1}$, (b) $N_c=5\times10^7 m^{-1}$, (c) $N_c=10^8 m^{-1}$, and (d) $N_c=15\times10^8 m^{-1}$.

In this work we have investigated the changes in performance of MOS CNTFET by changing the doping level of channel at different diameter values. It is found that for lower "n" values, an increase in the doping level causes the current ratio to decrease. But for higher "n" values, growth in channel doping level, up to a certain amount, leads to improvement in current ratio and after that point the overall effect of increasing the doping level is performance degradation. If the on-state current of device is supposed at 0.4V gate field, the results are approximately the same. So with respect to the results presented here, an optimum point of channel doping level is $5\times10^8$ $m^{-1}$ or more which leads to the maximum $I_{on}/I_{off}$ ratio for high "n" values, e.g. n=17 and n=19. Also a channel without doping is the best choice for gaining the high $I_{on}/I_{off}$ ratio for lower "n" values, e.g. n=10.